\documentclass[11pt,twocolumn]{article}

\usepackage{amsmath}
\usepackage{cite}
\usepackage{graphicx}
\usepackage{siunitx}
\usepackage{textcomp}
\usepackage{microtype}
\usepackage{subcaption}
\usepackage{multirow}
\usepackage{authblk}
\usepackage{fancyhdr}
\usepackage{url}
\usepackage{color,soul}

\title{Poissonian communications: free space optical data transfer at the few-photon level}

\author[1]{Alexander D. Griffiths}
\author[1]{Johannes Herrnsdorf}
\author[2]{Christopher Lowe}
\author[2]{Malcolm Macdonald}
\author[3]{Robert Henderson}
\author[1,*]{Michael J. Strain}
\author[1]{Martin D. Dawson}

\affil[1]{Institute of Photonics, University of Strathclyde, Glasgow}
\affil[2]{Department of Mechanical \& Aerospace Engineering, University of Strathclyde, Glasgow}
\affil[3]{CMOS Sensors \& Systems Group, University of Edinburgh, Edinburgh}

\affil[*]{Corresponding author: michael.strain@strath.ac.uk}

\begin{document}
\twocolumn[
\begin{@twocolumnfalse}
	\maketitle
\begin{abstract}
Communicating information at the few photon level typically requires some complexity in the transmitter or receiver in order to operate in the presence of noise.   This in turn incurs expense in the necessary spatial volume and power consumption of the system.  In this work we present a self-synchronised free-space optical communications system based on simple, compact and low power consumption semiconductor devices.  A temporal encoding method, implemented using a gallium nitride micro-LED source and a silicon single photon avalanche photo-detector (SPAD) demonstrates data transmission at rates up to 100~kb/s for 8.25~pW received power, corresponding to 27 photons per bit. Furthermore, the signals can be decoded in the presence of both constant and modulated background noise at levels significantly exceeding the signal power.  The system’s low power consumption and modest electronics requirements are demonstrated employing it as a communications channel between two nano-satellite simulator systems.\\
\end{abstract}

  \end{@twocolumnfalse}
]

\section{Introduction}

Conventional optical wireless communications (OWC) involves the modulation of the optical intensity of a light source, such as a light-emitting diode (LED) or laser, and direct detection of the output light \cite{Rajbhandari2016a}. When transmitting over long distances, or through high loss media, received power can become greatly reduced. Ultimately, the modulated signal can become lost in noise, arising from background light in the optical channel or within the receiver electronic components themselves. A number of schemes have been developed to operate using an exceptionally low number of received photons per bit, including error correction coding and higher order modulation schemes \cite{Boroson2007}, coherent detection at Mb/s \cite{Stevens2008} and Gb/s \cite{Geisler2013} data rates,  phase shift keying (PSK) and rate-1/2 forward error correction (FEC) codes. Such schemes are approaching the standard quantum limit (SQL) for sensitivity, and make use of complex components such as balanced photodetectors, local oscillators and optical phase-locked loops. It should also be noted that the SQL is not the fundamental quantum limit for optical communications, and can be surpassed by employing quantum receivers \cite{Chen2011,Tsujino2011,Becerra2013}. 

A potentially simpler way to attain high receiver sensitivity is to make use of single photon detection and counting \cite{Robinson2006,Willis2012,Chitnis2015,Li2015e,Zimmerman2017}. With the use of FEC codes and high order pulse position modulation (PPM) \cite{Alsolami2012}, photon counting systems can also operate with extremely low numbers of photons per bit \cite{Boroson2007}. However, these systems are critically dependent on temporal synchronisation of the transmitter and receiver. The high sensitivity of single photon counting techniques, combined with PPM and arrayed receivers has potential for deep-space communication links, operating at megabit rates. Such systems have been proposed for communications to Mars orbit \cite{Mendenhall2007}, and demonstrated for Lunar range optical communications \cite{Boroson2014}. These detection schemes are however limited by the optical noise in the channel and the dark count rate of the detector \cite{Liao2017}. 

Transmitters that generate coincident photon pairs may be used in time gating schemes that can reduce the effects of noise in the system \cite{Lekki2008,Hizlan2009}. Counting coincidences allows data to be sent with extremely low levels of received power over long distances \cite{Seward1991, Hong11}, and can be used for quantum key distribution \cite{Hughes2000,Gordon2004,Takenaka2017}. These methods typically require high efficiency photon pair sources with large form factor requirements on the internal and transceiver optics, and in some cases, secondary clock channels.

Here we demonstrate a novel optical transmission scheme, suitable for OWC at ultra-low light levels, requiring only a single, low photon flux channel. Compared with existing methods, this scheme is implemented with simple and widely available semiconductor components and electronics in a compact format and with low power consumption. Importantly, the method operates under the presence of both constant and modulated background noise, which is enabled by the encoding of data in the timing statistics of the received photons.

The data is formatted as temporally correlated optical pulses, in this case from an LED, transmitted over a single optical channel. The signal is detected with a single photon avalanche diode (SPAD), the output of which is a series of photon detection events including both data and noise sources. By calculating the autocorrelation of the SPAD output, discretised in a histogram, peaks can be observed corresponding to the specified temporal correlation of the received pulses. Data can therefore be encoded in the presence or absence of correlation peaks at predetermined time intervals. This allows undesired background signals to be rejected efficiently, unless they coincide with the data signal simultaneously on two separate timescales, namely the correlation time and the symbol rate. At data rates of a few kb/s, data can be transmitted with received signal power at a fraction of the background signal power. By trading off the data rate, the scheme can in principle adapt to arbitrarily low signal to noise ratio (SNR) as the link fidelity is governed by the Poissonian count statistics rather than conventional SNR. Furthermore, the experiments were performed using a simple optical and electronic setup comprising a GaN micro-LED transmitter and a single silicon SPAD receiver interfaced with field-programmable gate arrays (FPGAs). The scheme is also scalable to arrayed sources and receivers and can thus be extended to a multiple input multiple output configuration \cite{Rajbhandari2016a}. All this opens up potential for application in areas such as the internet of things, where a network of individual elements will need to communicate under low power, potentially powered through energy harvesting, with modest data rates. In this light, we demonstrate the suitability of the scheme for nano-satellite communications at such low power levels that transmission distances in the 10's of km range are feasible using a single LED emitter.

The following sections discuss the details of the transmission scheme, its current implementation, data transmission results and a demonstration of the system's suitability for inter-satellite communications, such as shown in Figure \ref{fig_Fig1}a.

\section{Results}
\subsection{Time correlation encoding scheme}

The transmission scheme presented here, inspired by time-correlated single photon counting (TCSPC) techniques often used for fluorescence lifetime imaging \cite{Becker2015}, involves the use of a single SPAD to receive time correlated signals at a single photon level. Analysis of the SPAD response to incoming light over an interval $[-t_1, t_1]$ shows that the correlation count density function $g(\tau)\mathrm{d}t'$ of recording two subsequent SPAD counts with temporal separation in the interval $[\tau, \tau+\mathrm{d}t']$ is given by:

\begin{equation}\label{eq_g}
g(\tau) = \int_{-t_1}^{t_1}\mathrm{d}t\;f(t)f(t+\tau).
\end{equation}

Here $f(t)$ is the temporal probability distribution of received SPAD pulses, which is determined by the optical signal from the transmitter. Full analysis is given in the supplementary material. If a suitable optical source transmits pulses with a time separation of $T$, $g(\tau)$ will show a peak at $\tau = T$, as the probability of observing SPAD pulses separated by $T$ is increased. Equation \ref{eq_g} is the autocorrelation of $f(t)$, so it is expected that peaks in $g(\tau)$ would have a width of $2t_{pulse}$, where $t_{pulse}$ is the width of the optical pulse. It is important that $T > \tau_{d}$, the dead time of the SPAD, as otherwise the SPAD would not recover from the first pulse in time to see the second.  This restriction can be lifted by making use of a SPAD array \cite{Fisher2013, Gnecchi2016}, however here we consider the use of only a single SPAD. The presence and/or temporal position of peaks in $g(\tau)$ directly depends on the sequence of optical pulses from the transmitter, and therefore can be used as a means of transmitting data.

In reality, the SPAD output is not a continuous probability distribution, but a series of discrete photon detection events. These events can occur due to the optical pulses from the transmitter, background photons, or dark counts. The SPAD output signal will be sampled over a time period, into time bins $t_{i}$, $i=1,\ldots,N_{s}$, where $N_{s}$ is the total number of time bins. Each bin contains a number of counts $f_{i}$. If these bins are chosen to be smaller than $\tau_{d}$,  $f_{i}$ will only have values of 0 or 1. Correlation time will also be discretised into $\tau_{j}$, $j=1,\ldots,N_{\tau}$. For a single pair of pulses a correlation either is or is not detected. This single correlation is indistinguishable from a random correlation of counts, so the optical signal must be repeated many times to produce a usable histogram of correlation counts. If correlation time bin size is chosen as an integer multiple of sampling bin size, $\tau_{bin} = kt_{bin}$, we can define start and stop indices for correlating across $i$ as:

\begin{align}
n_{start} &= \frac{\tau_1}{t_{bin}} \label{eq:n_{start}},\\
n_{stop} &= n_{start} + kN_\tau - 1 \label{eq:n_{stop}}.
\end{align}

With this, the discrete form of Equation \ref{eq_g} is:

\begin{equation}\label{eq_gfinal}
g(\tau_j) = \sum_{i=1}^{N_s-n_{stop}}\sum_{l=0}^{k-1}f_if_{i+n_{start}+(j-1)k+l}.
\end{equation}

As $f_{i}$ is a binary value, and the output from the SPAD is a transistor-transistor logic (TTL) signal, the summation could be implemented with simple logic circuits. 

Encoding data in $g(\tau_{j})$ has the potential to allow data transfer at exceptionally low light levels, and in the presence of significant background illumination. In order to detect correlations, the receiver requires the detection of a single photon from each optical pulse. Such conditions allow average received power to be extremely low, in the range of \SI{}{\pico\watt}. The trade-off in this transmission scheme is that the data rate is expected to be relatively modest, as the optical signal must be repeated several times in order to generate a distinguishable signal in $g(\tau_{j})$.

There are several potential ways to encode data in $g(\tau_{j})$. Forms of pulse position modulation (PPM) or pulse amplitude modulation (PAM) are discussed in the supplementary material. Here we consider the simplest form of encoding, on-off keying (OOK), where data can be encoded using a single pulse time separation. On transmission of the symbol `1', pulses of width $t_{pulse}$ are transmitted continuously with a fixed time separation $T$, so $g(\tau_{j})$ will show a peak at $\tau = T$. On transmission of the symbol `0' no pulses are transmitted, producing only background peaks in $g(\tau_{j})$.

\begin{figure*}
	\centering
	\includegraphics[width = \textwidth]{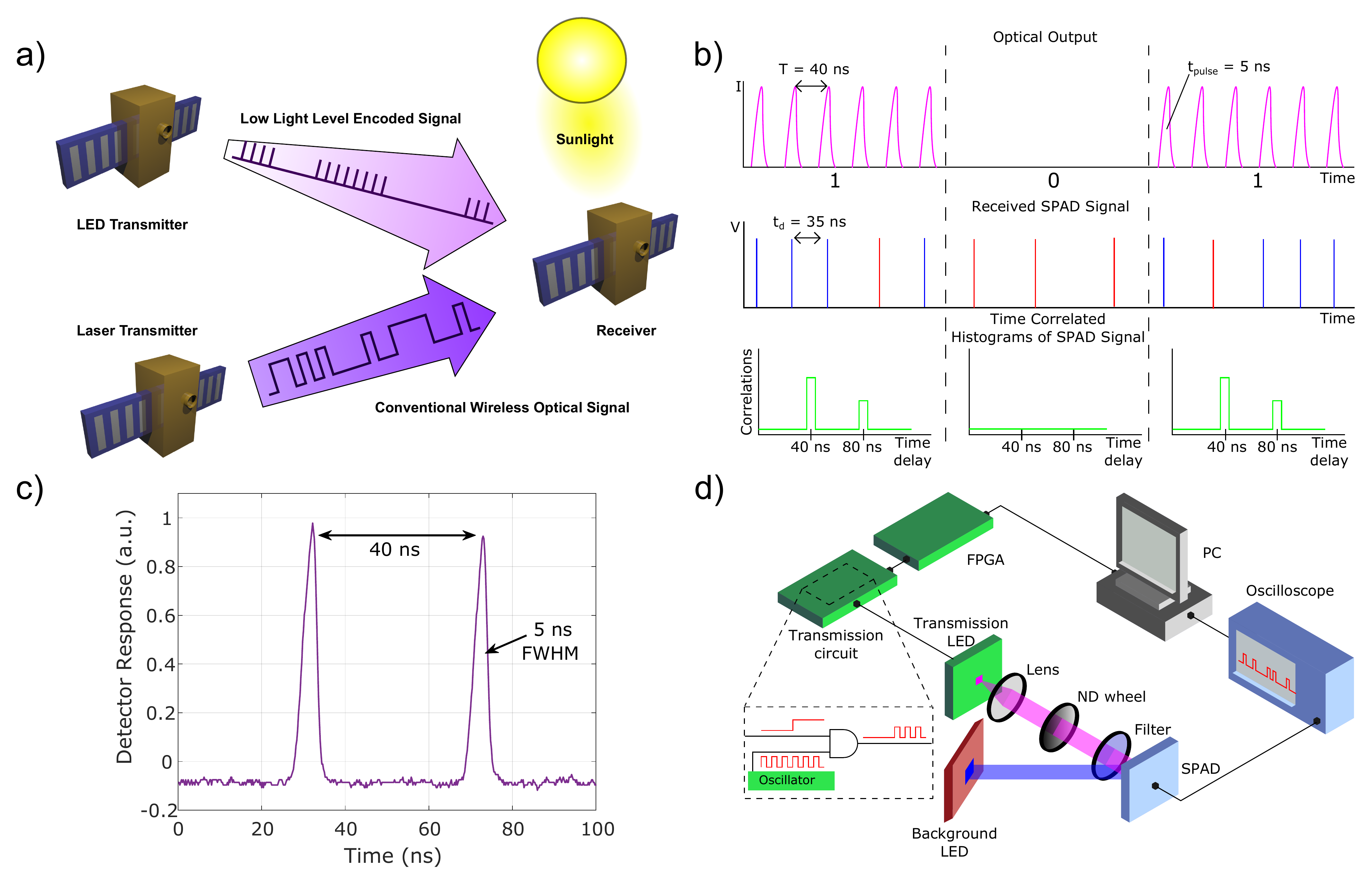}
	\caption{a) Application scenario for inter-satellite communications: The low light level signal encodes data in trains of time-correlated optical pulses and is compatible for use with low power and divergent LED emitters. It is resilient against DC background such as sunlight and also insensitive to most kinds of AC background such as conventional wireless optical signals. b) Schematic of the transmission scheme used. The upper plot shows the LED output on transmission of `0' and `1'. The middle plot shows the SPAD response to the LED signal. The lower plots show the calculated correlation histograms for each data interval.  c) Measured pulse pair from a micro-LED source. d) Schematic of the experimental setup.}
	\label{fig_Fig1}
\end{figure*}

A schematic of the expected waveforms is shown in Figure \ref{fig_Fig1}b. In this example, optical pulses are sent with time separation of $T = 40$ ns. If we choose $t_{pulse} = 5$ ns, deliberately less than $\tau_{d}$, then only one photon can possibly be detected from each pulse, indicated by the blue SPAD signals in Figure \ref{fig_Fig1}b. In reality the detection rate will be less than one per pulse, and pulses can also be missed if received during the dead time after a noise pulse, indicated in red. Time correlation of the measured events from the SPAD is performed over a data interval, producing a histogram with peaks at \SI{40}{\nano\second} intervals for transmission of a `1', and a background correlation level for transmission of a `0', with the noise floor determined by ambient background light and detector dark count rate. Simply applying a threshold to the histogram bin generated for each symbol at a delay \SI{40}{\nano\second} allows decoding of a binary stream. This threshold will have to be sufficient to reject correlation counts from background and dark count correlations.

Finally, a crucial feature of this method is that it is robust to temporal jitter between the transmitter and receiver, unlike other forms of photon counting schemes.  Synchronisation of the system can be easily achieved by using an embedded clock in the transmitted data, as discussed in the supplementary material.

\subsection{Experimental demonstration}

The scheme detailed above was realised using a GaN violet emitting (405 nm), micro-LED device as the transmitter and a~ silicon Single Photon Avalanche Detector (SPAD) as the receiver. The LED chip was bonded to a custom CMOS driver allowing short pulse operation, with durations of 5~ns. Data signal modulation was applied as a slower on-off keying of the short pulse train. Figure \ref{fig_Fig1}c shows a measured pair of pulses from the micro-LED of 5 ns duration and with a relative delay of 40~ns. A variable neutral density filter was placed between the emitter and detector to control the received power at the SPAD. A schematic of the measurement setup is shown in Figure 1D. Full details of the devices and electronic drivers are given in the methods section.

\subsection{Signal-to-noise ratio}

To set an operation threshold for the system, a figure of merit similar to the classical signal to noise ratio must be defined.  In this method, it is the distinguishability of the correlation peak in the $g(\tau)$ function that indicates the robustness of the classical information recovery to noise. Conventional SNR can be defined as in Equation \ref{eq_SNR}, where $N_{signal}$ is average signal correlation counts and $N_{noise}$ is average noise correlation counts:

\begin{equation}\label{eq_SNR}
SNR = 10\log_{10}\frac{N_{signal}}{N_{noise}}.
\end{equation}

As the number of pulse repetitions increases, the correlation counting interval increases, causing both $N_{signal}$ and $N_{noise}$ to increase at linear rates. This results in a constant SNR, which does not reflect the observed increase in distinguishability of signal correlations with increasing pulse repetitions.

\begin{figure*}
	\centering
	\includegraphics[width = \textwidth]{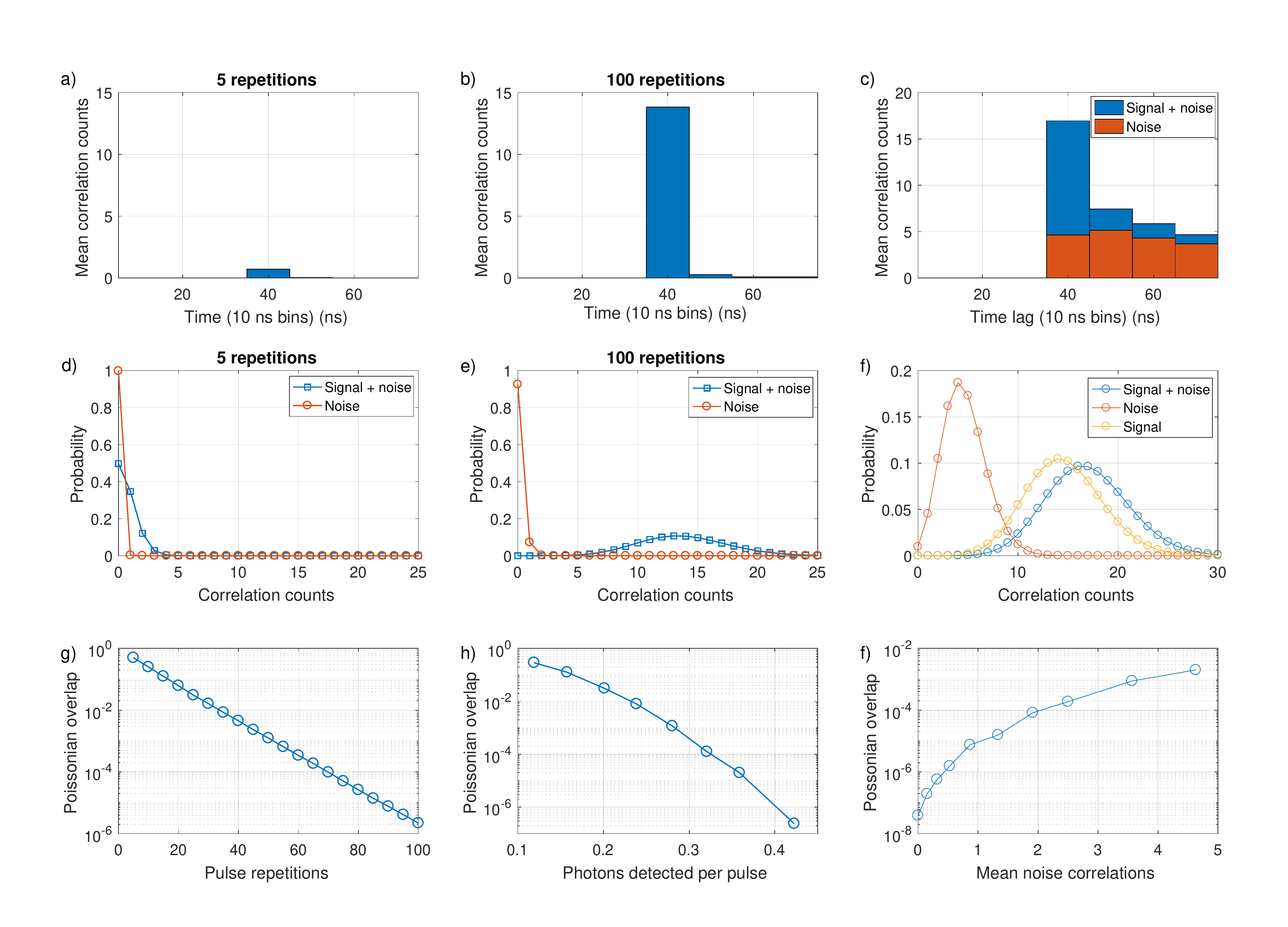}
	\caption{Measured correlation histograms for a) 5, b) 100 pulse pair cycle repetitions and c) 100 repetitions with background noise. Poissonian distributions for the signal and noise correlation counts for d) 5, e) 100 pulse repetitions and f) 100 repetitions with background noise. g--i) Poissonian overlap according to Equation~\ref{eq_Overlap} as a function of g) number of pulse repetitions, h) average number of photons detected per pulse, and i) average number of background noise correlations.}
	\label{fig_Fig2}
\end{figure*}

Instead, it is more useful to consider the statistical distribution of correlation counts for signal and noise. Photon counting experiments were undertaken using the experimental setup described above, with a received power at the detector of 38~pW, corresponding to a detector count rate of $1.07\times 10^7$~Hz, in a dark lab environment with an average background count rate of 619~Hz. Note that this count rate contains both dark counts within the detector and counts from the small amount of ambient light. The delay correlations of detected photons were binned with a resolution of 10~ns, where the transmitted pulse delay was set at 40~ns. Figure \ref{fig_Fig2}a and b show average histograms of received photon correlations for 5 and 100 pulse pair repetitions, respectively. Figure \ref{fig_Fig2}c shows the histogram for 100 pulse pair repetitions under high background conditions, displaying the correlation histogram due to background noise alone, and signal with noise. The background count rate for this measurement was $10^7$~Hz.

In Figure \ref{fig_Fig2}a and b, the signal is defined as the number of correlations in the 40~ns delay time bin, and the noise correlation count is taken from the 60~ns delay bin. Correlation counts should follow a Poissonian distribution, as they are discrete independent events. Figure \ref{fig_Fig2}d and e are the measured Poissonian distributions for signal with noise and noise alone correlation counts at 5 and 100 repetitions of the dual pulse cycle, respectively, taken from 1500 independent measurements of each case. 

At 5 repetitions, the probability distributions for signal and noise are strongly overlapped. Thus a correlation count peak due to signal transmission is difficult to distinguish from a correlation count peak due to random background and dark counts. At 100 repetitions, the overlap of signal and noise distributions is significantly reduced,  making distinction much easier. A histogram threshold equates to a point along the x-axis of the distribution plots. Evidently a threshold point of 2 at 5 repetitions would result in many signal correlation peaks being rejected as noise, whereas at 100 repetitions the majority of signal correlation peaks would be correctly identified, and noise correlation peaks rejected.

Under high background noise conditions, the number of correlations from noise is increased. Figure \ref{fig_Fig2}f shows the Poissonian distributions for the background noise, signal alone (identical to Figure \ref{fig_Fig2}e) and the combined signal and noise. To properly decode received signals, the threshold must distinguish between the correlations due to noise, from those due to the signal, as both will be present on transmission of `1'.

Distinguishability is therefore determined by the overlap of the Poisson distributions for (i) the total signal and noise contributions, $P_{T}(k)$, and (ii) the noise alone $P_{n}(k)$. Here, $P_{n}(k)$ is the the probability of $k$ correlation counts occurring due to noise with mean $\lambda$, given by Equation~\ref{eq_Poisson}. Similarly, the distribution of correlation counts $P_{s}(k)$ when a signal is present without noise is also given by Equation \ref{eq_Poisson}. The distribution of correlation counts in the presence of both signal and noise, $P_{T}(k)$ is related to $P_{s}(k)$ and $P_{n}(k)$ via Equation \ref{eq_spread}. Figures \ref{fig_Fig2}g--i show the calculated overlap of these distributions as defined in Equation \ref{eq_Overlap}:

\begin{align}\label{eq_Overlap}
Overlap = \sum_{k=0}^{\infty}P_{T}(k)P_{n}(k),  \\
P(k) = \frac{\lambda^{k}e^{-\lambda}}{k!}\label{eq_Poisson}\\
P_{T}(k) = \sum_{m=0}^{k}P_n(m)P_{s}(k-m).\label{eq_spread}
\end{align}

The overlap of the probability distributions reduces exponentially with increasing pulse repetitions. The rate of decay of the overlap will depend on the received signal power. Figure \ref{fig_Fig2}g is valid for a received power of \SI{38}{\pico\watt}. Higher received power would result in a steeper decay, and lower power in a more shallow decay.

In addition to the number of repetitions, the Poisson distribution overlap will depend on the number of photons detected from each pulse. As the pulses used are shorter than the dead time of the SPAD, the maximum number of photons that can be detected is one. Figure \ref{fig_Fig2}h shows the overlap for 100 pulse repetitions as a function of received photons per pulse. It can be seen that the overlap reduces faster than exponential. This is understood by noting that $\lambda$ in Equation~\ref{eq_Poisson} follows $\lambda = p_{ph}^{2}N_{rep}$, where $p_{ph}$ is the probability of detecting a photon from a single pulse, and $N_{rep}$ is the number of pulse repetitions. This relationship gives rise to the linear appearance of Figure \ref{fig_Fig2}g and the quadratic appearance of Figure \ref{fig_Fig2}h.

The rate of random correlation counts due to the background noise will also affect the overlap of Poisson distributions, as increased noise correlations changes the distribution $P_n$. Figure \ref{fig_Fig2}i shows the calculated overlap as background correlation counts increases. Larger numbers of noise correlations increase the overlap between $P_n$ and $P_{T}$ with a sub-exponential trend.

Therefore, the distinguishability of binary 0 and 1 is governed by the Poissonian overlap, Equation~\ref{eq_Overlap}, and in turn depends on the number of sampled pulse repetitions, the received signal power, and the background intensity, with the first two parameters dominating.

\subsection{Data rates}

The achievable data rate of this system is determined by the number of pulse repetitions required to distinguish the signal, and hence the received power and the time separation between pulses. The SPAD response imposes a lower limit on this separation, due to the dead time $\tau_{d}$ and pulse width, $\tau_{pulse}$, giving an achievable data rate of:

\begin{equation}\label{eq_data}
R_{data} = \frac{1}{N_{rep}(\tau_{pulse}+\tau_{d})}.
\end{equation}

\begin{figure*}
	\centering
	\includegraphics[width = \textwidth]{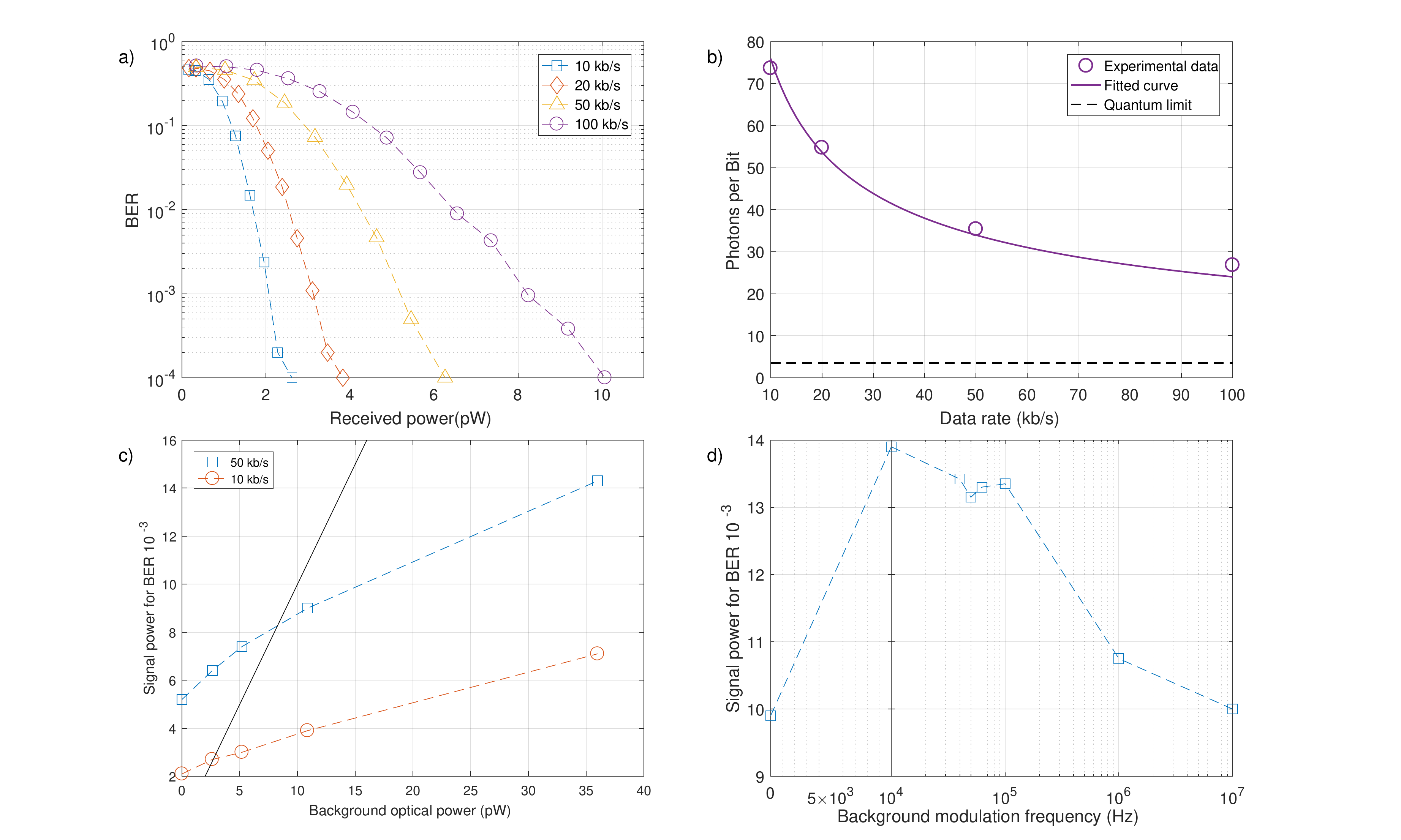}
	\caption{a) BER as a function of received signal power for varying data rates. b) Received photons per bit required to achieve a BER of less than $10^{-3}$ for varying data rates, fitted with a $x^{-\frac{1}{2}}$ relationship. The standard quantum limit for uncoded OOK at this BER is also shown. c) Required signal power to attain a BER of $10^{-3}$ under constant background power for 50 and 10~kb/s. Detailed conditions are shown in Table \ref{tab_BG}. Equal signal and background power is indicated by the solid line. d) Required signal power to attain a BER of $10^{-3}$ under modulated background conditions. Required power increases when background modulation rates are comparable to the signal data rate.}
	\label{fig_Fig3}
\end{figure*}

Where $N_{rep}$ is the number of pulse repetitions required to see a distinguishable peak in the correlation histogram. Use of a SPAD array could lift the restrictions imposed by dead time through pulse combining techniques \cite{Gnecchi2016}.  To demonstrate the system performance as a function of received power and data rate, bit error ratio (BER) measurements were made over these parameters. A pseudo-random bit sequence (PRBS) of $10^{4}$ bits was used, limited by the data processing capabilities of the oscilloscope and PC components in the measurement setup.  The ND filter wheel allowed control of received power, or equivalently, photon detection probability. Figure \ref{fig_Fig3}a shows BER curves for varying data transmission rates, taken with minimal background light. 

As detailed above, detection of a noise correlation is extremely unlikely. On transmission of a `0', no pulses are sent, meaning only noise correlations will be present. Bit errors arise almost exclusively from missed correlation counts on transmission of a `1' rather than noise correlations on transmission of `0'. Therefore, a decision threshold of 2 counts applied to the correlation histograms provides the lowest BER values.  At \SI{8.25}{\pico\watt} of average power, a data rate of 100 kb/s was possible with a BER of less than $10^{-3}$. Received optical power can be reduced at the expense of data rate. A data rate of 10~kb/s can be achieved at the same BER with \SI{2}{\pico\watt}. 

The power measurements quoted here and used in Figure \ref{fig_Fig3} are the incident optical power on the active area of the SPAD. This value has been calculated through numerical methods from the average count rate detected by the SPAD, details of the method can be found in the supplementary material. Detector count rate is the parameter which governs BER performance, however the optical power to attain the required counts will be influenced by the performance of the SPAD. Most importantly, the detection efficiency of the SPAD at 405~nm is 18\%, meaning the incident photon flux is higher than the detector count rate. More efficient photon detection would improve BER performance in terms of required power.

The system performance can also be described in terms of the number of received photons per bit. Figure \ref{fig_Fig3}b shows detected photons per bit for each data rate, at the level required for a BER of less than $10^{-3}$. The fitted curve is calculated from the relationship between correlation counts, received power and data rate. The number of signal correlations depends on the square of received power and is inversely proportional to the data rate $R_{data}$. In order to reach a given target BER, a certain constant number of signal correlation counts must be reached, meaning $(ph/s)^{2} \propto R_{data}$. As photons per bit is simply the required photons per second divided by data rate, we obtain Equation \ref{eq_fit}:

\begin{equation}\label{eq_fit}
ph/bit \propto \frac{1}{\sqrt{R_{data}}}.
\end{equation}

The 100 kb/s link is transmitting each bit with an average of 27 detected photons. This is relatively close to the standard quantum limit (SQL) for uncoded OOK, set by Poissonian photon statistics \cite{Shieh2010}. The SQL for a given BER is defined by the minimum number of photons required to distinguish a signal from noise. For a BER of $10^{-3}$, a minimum of 7 photons is required, to detect a `1'. Therefore an average of 3.5 photons per bit is required, assuming the probability of transmitting `0' or `1' is equal. The implemented scheme will be unable to reach the SQL, due to the correlation approach. Two photons are required for a single correlation detection, which itself has a Poissonian distribution that must be distinguished from noise.

Note that no FEC has been implemented here, so there is potential for improved efficiency. Furthermore only OOK style signalling has been implemented. Further efficiency improvements will be possible through use of PPM style transmission, to transmit multiple bits per correlation peak.

After correcting for detector efficiency, 27 detected photons equates to \SI{7.37e-17}{\joule} incident on the detector per bit. This exceptionally low energy demonstrates the suitability of the transmission scheme in low power or high loss systems. As mentioned above, more efficient photon detection would allow further reductions in energy received per bit.

\subsection{Robustness to noise}

A major advantage of this transmission scheme is that it is expected to be robust against background counts, as ambient light is generally uncorrelated on the time scale of 10s of ns. To verify this, BER curves were taken for increasing levels of background light using a secondary light source, as shown in Figure \ref{fig_Fig1}d. As background counts increase, the probability of detecting noise correlations increases. The threshold applied to the correlation histogram must then be increased to avoid erroneous detection of `0' symbols. This means higher received average power is required to obtain error-free transmission.

The results for varying background at data rates of 50 and 10~kb/s are shown in Figure \ref{fig_Fig3}c, with the background conditions detailed in Table \ref{tab_BG}. Full BER curves can be found in the supplementary material. $R_{bg}$ is the detector count rate from the background light alone, before the signal is sent. Background incident photon flux ($\Phi_{bg}$) is the number of photons incident on the detector, calculated using numerical methods described in the supplementary material. Optical power ($P_{bg}$) is then simply the photon flux multiplied by photon energy. Table \ref{tab_BG} also shows the detector count rate ($R_s$), incident photon flux ($\Phi_{inc}$) and optical power ($P_{inc}$) of the signal required to obtain a BER of $10^{-3}$ at 50~kb/s. 


\begin{table*}[th]
	\centering
	\caption{Table of conditions for BER curves in Figure \ref{fig_Fig3}c. Detector count rate ($R$), incident photon flux ($\Phi$) and optical power ($P$) are shown for background conditions (subscript bg) and for the signal at a BER of $10^{-3}$ (subscript s/inc).}
	\label{tab_BG}
	\begin{tabular}{cccccccc}
		\hline
		Curve & \begin{tabular}{c}$R_{bg}$ \\ ($s^{-1}$)\end{tabular} & \begin{tabular}{c}$\Phi_{bg}$\\ ($ph/s$) \end{tabular} & \begin{tabular}{c}$P_{bg}$\\($pW$)\end{tabular}  & \begin{tabular}{c}$R_s$ \\ ($s^{-1}$)\end{tabular} & \begin{tabular}{c}$\Phi_{inc}$\\ ($ph/s$) \end{tabular}& \begin{tabular}{c}$P_{inc}$\\($pW$)\end{tabular}  \\ \hline
		a	&	$619$ & \SI{2.95e3} & $1.45\times10^{-3}$ &	\SI{2.04e6} & \SI{1.06e7} & 5.2	\\
		b	&	$1.08\times10^{6}$	&  \SI{5.34e6} & 2.62 &	\SI{2.46e6}	&  \SI{1.30e7} & 6.4 \\
		c	&	$2.06\times10^{6}$	& \SI{1.06e7} & 5.19	&	\SI{2.80e6}	& \SI{1.51e7} & 7.4	\\
		d	&	$4.00\times10^{6}$	& \SI{2.21e7} & 10.87 &	\SI{3.29e6}	& \SI{1.81e7} & 8.9	\\
		e	&	$1.00\times10^{7}$	& \SI{7.33e7} & 35.98 &	\SI{4.84e6}	& \SI{2.91e7} & 14.3	\\ \hline
		
	\end{tabular}
	
\end{table*}

Figure \ref{fig_Fig3}c shows increasing levels of background optical power increases the signal power required to obtain a BER of less than $10^{-3}$. However, the power requirements are still very low. At high background levels, the required signal power is significantly lower than the power received from background illumination. With a background count rate of $10^7$, corresponding to an optical power of 35.98~pW, less than 15~pW of signal power is required for a 50~kb/s link, and less than 8~pW for 10~kb/s. This background illumination level is somewhat extreme, as the detector will saturate at \SI{1.4e7}, and normal room lighting in this setup gives a background count rate of \SI{2.33e5}. In a practical system, background counts can be minimised using suitable optical filters.

The system was also measured under modulated background illumination.  Since the number of detected correlation counts depends on the square of received power, a high modulation rate background should interfere in the same manner as a DC signal at the root-mean-squared (RMS) of its count rate. For this reason, background signals used in the following experiments maintain similar RMS photon count rates for comparison to DC measurements. The power required to maintain a BER of $10^{-3}$ is shown in Figure \ref{fig_Fig3}d, with the background conditions and full BER curves shown in the supplementary material. The RMS background optical power was approximately 15~pW for all measurements.

Figure \ref{fig_Fig3}d shows two distinct groups of results. The high background modulation rates of 1 and 10~MHz show similar required signal power to constant background conditions. This can be attributed to 2 factors. Firstly, the background signal completes many cycles within a single bit period of the correlation link, making its modulation less significant. Secondly, the dead time of the SPAD restricts the number of photons that can be detected per background cycle, reducing the difference between a high and low level, further making the background signal act like constant interference. When the background modulation rate is close to the correlation link data rate, the BER performance is degraded, requiring approximately 40\% more received power. However, all conditions still reach a BER of less than $10^{-3}$ for less than 14~pW of received signal power. This reduction in performance occurs as the background signal is now generating different levels of noise correlations from one bit period to the next. This makes it more difficult to choose a suitable correlation threshold, and increases the BER.

Under all background conditions, the signal is transmitted with a lower photon count rate than the background signal, demonstrating low power performance even with high power modulated background interference. 

\subsection{Satellite systems demonstration}

The communications system presented here is applicable in many scenarios, but is particularly attractive for inter-satellite links.  The semiconductor devices are extremely compact, low power consumption and readily integrated with control electronics. LED based visible light communications shows potential for use with cube satellites \cite{Kawamura2013}. The robustness of the signal to background noise and operation at ultra-low power levels, means that it could be implemented without the high accuracy pointing requirements and telescope optics of current satellite systems.

\begin{figure*}
	\centering
	\includegraphics[width = \textwidth]{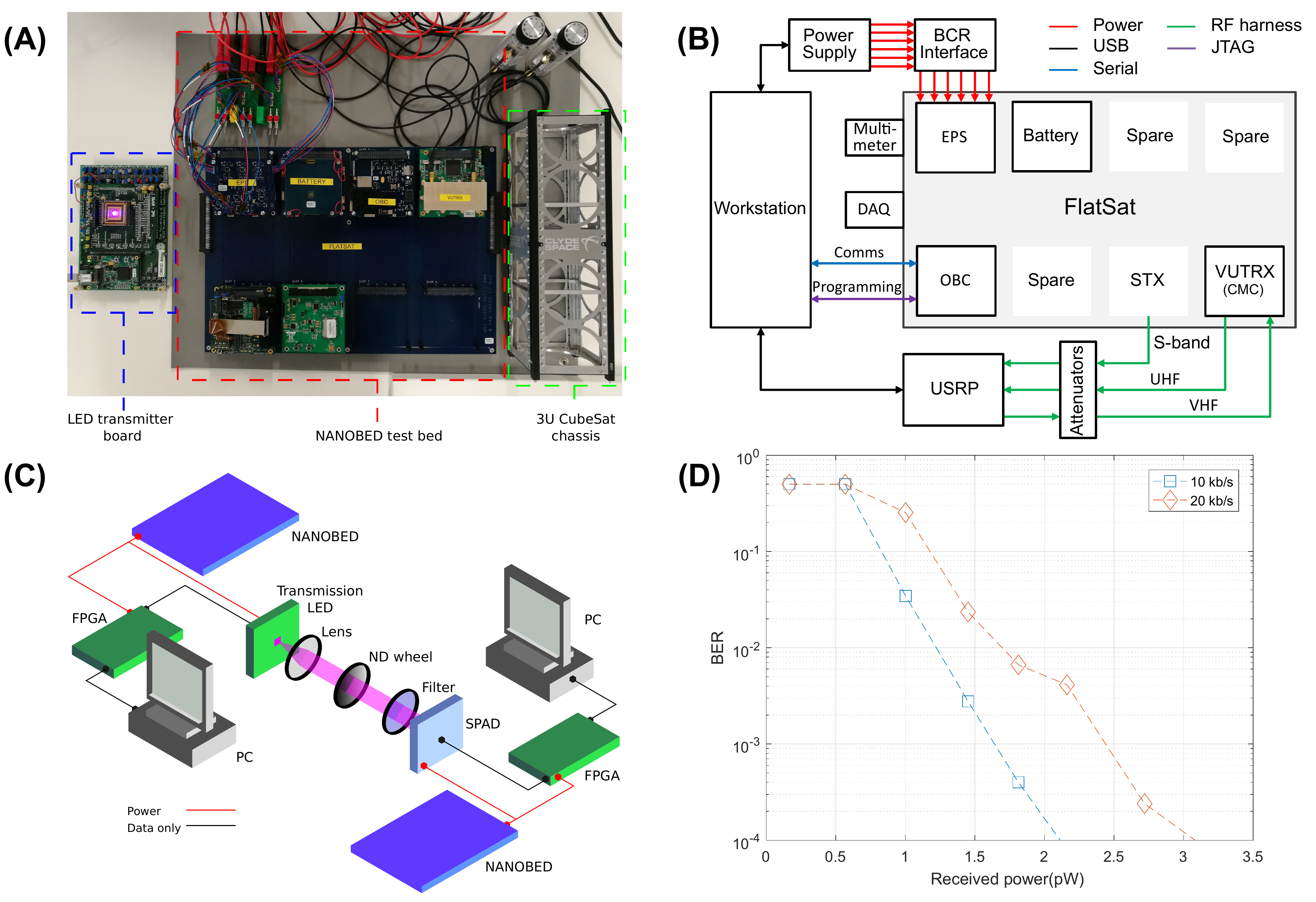}
	\caption{a) Photograph of the micro-LED transmitter board (left), NANOBED ``Flatsat" test bed (centre) and a standard 3U CubeSat chassis (right).  b)  Schematic of the NANOBED architecture. The configuration allows access to individual components of the CubeSat, including the on-board computer (OBC) and electrical power supply (EPS). Radio links are provided by an S-band transmitter (STX) and VHF/UHF Transceiver (VUTRX) connected to a universal software radio peripheral (USRP). c) Schematic of the experimental setup for real time data transmission. d) BER performance of the communication link under power from the NANOBED system.}
	\label{fig_Fig4}
\end{figure*}

To highlight this capability the system was tested in the nano-satellite mission development environment, NANOBED, shown in Figure \ref{fig_Fig4}a. The NANOBED is a nano-satellite hardware and software test-bed, with mission planning facilities, details of which are given in the methods section.  A block diagram of the NANOBED setup is presented in Figure \ref{fig_Fig4}b. Furthermore, to demonstrate that the full system functionality was able to be powered by the NANOBED, a real time decoder, incorporating embedded clock signal recovery, was implemented on a FPGA platform to replace the oscilloscope and PC components in the characterisation setup. Details of this setup, shown in Figure \ref{fig_Fig4}c, are given in the methods section. Details of the integrated clock recovery are given in the supplementary material.

The LED transmission system was integrated with one NANOBED system, while the SPAD receiver system was integrated with a second.  In this work, the NANOBED solar panel emulating power sources are used to supply the transmitter and receiver devices via the nano-satellite electrical power supply (EPS) and battery units, simulating an in-orbit scenario. The transmitter side of the real time link requires a single FPGA board, from which the CMOS micro-LED array is powered and controlled. On the receiver side, the commercial SPAD module requires a 6~V DC supply, and a second FPGA is used to process the received signals. A summary of typical power consumption is shown in Table \ref{tab_Power}.

\begin{table*}[th]
	\centering
	\caption{Typical power requirements of the communication system}
	\label{tab_Power}
	\begin{tabular}{cccc}
		\hline
		&Voltage (V) & Current (mA) & Power (W) \\ \hline
		Transmitter & 5	& 181 & 0.905	\\
		Receiver & 5 & 122 & 0.610  \\
		& 6 & 595 & 3.570 \\ \hline
	\end{tabular}
	
\end{table*}

For the laboratory demonstration the transmitter and receiver were placed 4~m apart with the micro-LED pixel projecting the light across a 4~cm wide square with received power controlled using a neutral density filter.   A micro-LED emitter at 450~nm was used to improve the Photon Detection Probability (PDP) to 25~\%.  As shown in Figure \ref{fig_Fig4}d, the live link requires 2.5~pW of received power to maintain a BER of $10^{-3}$ at 20 kb/s. On a \SI{20}{\micro\meter} diameter SPAD, 3~pW corresponds to an intensity of \SI{9.5}{\milli\watt\per\meter\squared}. To provide this over the projected 4~cm wide square, the micro-LED must emit an average power of only \SI{15.3}{\micro\watt}.

\section{Discussion}

We have demonstrated a transmission scheme suitable for ultra-low light level optical wireless communications. By transmitting temporally correlated signals, data communications can be performed at extremely low light levels, with received power on the order of pico-Watts. Signals can be transmitted using an LED, and received with a single SPAD. A 100~kb/s link has been achieved with a BER of less than $10^{-3}$ at a received power of 8.25~pW, close to the standard quantum limit for uncoded OOK, while received power can be reduced at the expense of data rate.

The scheme is robust to background light, with only a minor increase in required power for very high background conditions. Modulated background signals appear to have little additional influence over that of continuous background, suggesting the scheme could be used in parallel with other optical communications with minimal interference. Furthermore, multiple transmission systems using this scheme could operate without interfering with each other, simply by using different pulse time separations.

A real time transmission setup has been demonstrated, showing a method for clock synchronisation and determination of a threshold level. The current, unoptimised implementation allows a data rate up to 20~kb/s, with only a minor reduction in performance when compared to offline processed transmission. In addition, the real time transmission link has been demonstrated in a simulated satellite environment, at a received power density of 9.5~mW/m2. For an example inter-satellite link distance of 40~km, an LED average power emission of 1~mW and a detector optical aperture of 10~mm, the required pointing accuracy corresponding at this power density, or half angle divergence of the source, would be $\approx 0.1~$degree, which represents a significant relaxation in orientation tolerance with respect to current laser based systems. Additionally, GaN LEDs at low current densities show higher wall-plug efficiencies than laser diode counterparts \cite{Piprek2016}, further enhancing the power consumption characteristics of the system.

The modest data rates presented in this work are dominated by two key factors, firstly the requirement of this protocol for correlating many repetitions of a pulse pattern, and secondly, the dead time of the SPAD detector itself.  We also demonstrated the simplest transmission protocol case of on-off keying, with no error correction codes applied. 

Data rate and photons per bit efficiency can both be improved through relatively straightforward modifications to the system.  By using a SPAD array as a receiver rather than a single device, the dead time limitation can be overcome and therefore higher data rates achieved.  In addition by implementing a form of pulse position modulation, and implementing powerful FEC codes, the photons per bit transmission efficiency can be improved. Finally, data rates may be enhanced by using a form of pulse amplitude modulation, however the received power requirement would also increase. 

This transmission protocol has clear applications in communications systems for long range or high loss environments, but is also equally applicable in microscopy or low light level imaging systems when coupled with a SPAD imaging array, and can be implemented using a wide range of pulsed optical sources dependent on the application.

\section{Methods}

\subsection{Optical transmitter and receiver realisation}
The transmitter used for the results presented here is a complementary metal oxide semiconductor (CMOS) integrated gallium nitride micro-LED pixel. Details and fabrication of comparable devices can be found in \cite{Zhang2013}. The micro-LED pixel is a square $100 \times 100$ \SI{}{\micro\meter} in size, and part of a $16 \times 16$ array with a \SI{405}{\nano\meter} emission wavelength.

The micro-LED array is fabricated in flip chip format, and bump-bonded onto CMOS control electronics. The CMOS electronics allow the LEDs to be modulated in a pulsed mode, triggered by the falling edge of an input logic signal. The shortest stable optical pulses $t_{pulse}$ that could be generated with this device and control system were \SI{5}{\nano\second}. 

In order to produce pulses for the OOK transmission, a data signal produced by a field programmable gate array (FPGA) is sent to a simple transmission circuit. Here the data signal is combined with an oscillator producing a signal of square waves with a period of \SI{40}{\nano\second}, through an AND gate, as shown in Figure \ref{fig_Fig1}d.

The SPAD receiver is a commercial module (Thorlabs SPCM20A), with a detector active area diameter of \SI{20}{\micro\meter}. The dead time of the detector is \SI{35}{\nano\second}, and the typical dark count rate is 25 Hz. At \SI{405}{\nano\meter} and \SI{450}{\nano\meter} the photon detection probability (PDP) is 18\% and 25\% respectively. The module outputs \SI{3}{\volt} logic signals indicating photon counts. This signal is sent to an oscilloscope, and collected by the PC for offline processing of $g(\tau_{j})$. In a practical system, this processing could be performed by digital logic circuits

The LED output is collimated with a lens (Thorlabs C220TME-A) and transmitted through a graded neutral density (ND) wheel (Thorlabs NDC-50C-4M-A). A \SI{450}{\nano\meter} shortpass filter is used in front of the SPAD to reject additional background light. This filter is removed for the experiments assessing performance under high background conditions. The pixel is imaged onto the SPAD active area. As the pixel image is approximately a \SI{7}{\milli\meter} square, only a small portion of the light is imaged on to the SPAD circular active area of diameter \SI{20}{\micro\meter}. In a practical system, receiver optics could be used to collect more light on to the active area of the SPAD, allowing higher degrees of loss through the channel.

Received optical power is calculated numerically from the average number of photon counts detected. This method accounts for detector dead time, and photon detection probability at the operational wavelength. Details of the calculation can be found in the supplementary material.

To assess the effects of DC background illumination a commercial \SI{450}{\nano\meter} LED (OSRAM LD CQ7P) was placed within a few centimetres of the transmitter LED, directed towards the SPAD, as shown in the setup schematic in Figure \ref{fig_Fig1}d. By increasing the driving current for the commercial LED, the background counts could be controlled. 

The modulated background optical signal was generated using a commercial \SI{450}{\nano\meter} LED (OSRAM LERTDUW S2W) modulated with a transistor. This commercial LED had a modulation bandwidth of 15.9~MHz, and was placed within a few centimetres of the transmitting LED. Modulating this LED with a PRBS effectively simulates operation of the correlation link in an environment with conventional optical wireless communication links. 

\subsection{Real Time Link}
\label{sec_RealTime}
The results presented in the first part of the manuscript are all obtained with offline decoding. The output from the SPAD was collected with an oscilloscope, and the traces repeated many times to build a data sequence long enough for BER analysis. In addition, the oscilloscope was triggered from the transmitting FPGA, bypassing the issue of transmitter and receiver synchronisation. To demonstrate a more practical system, a real-time transmitter and receiver have been developed using FPGAs.

A synchronisation system has been implemented, involving data transmission in frames consisting of a 6 bit clock word and 32 data bits. The carefully chosen clock word, `001101', allows both frame level and symbol level synchronisation of data streams. Details on choice of clock work and synchronisation methods can be found in the supplementary material. A block diagram of the experimental setup for real-time transmission is shown in Figure \ref{fig_Fig4}c. On the transmitter side, the FPGA is used to generate a data stream in frames, with the 6 bit clock word. In contrast to the offline setup in Figure \ref{fig_Fig1}d, the FPGA now directly supplies the falling edge trigger for the LED board, without the need for extra logic circuitry. The receiver FPGA is connected to a separate PC, and clock synchronisation removes the need for a trigger from the transmitter. However, due to limitations from the FPGA boards, the achievable data rates with the real-time setup are limited to 20~kb/s. It should also be noted that the data rates quoted here include transmission of the clock word. This 18.75\% overhead reduces useful data transfer to 8.42 and 16.84~kb/s for 10 and 20~kb/s links respectively.

\subsection{NANOBED Satellite Simulator Experiments}
The LED transmitter and SPAD receiver systems were independently powered by separate NANOBED systems, positioned approximately 4~m apart. A 450~nm micro-LED was used, focussed on the receiver plane using an 8~mm focal length lens (Thorlabs C240TME-A), giving a pixel image size at of approximately 4~cm. To increase received power on the \SI{20}{\micro\meter} diameter SPAD, a 35~mm focal length collection lens (Thorlabs ACL4532U-A) was used. 

The satellite simulator test bed is a FlatSat-configured CubeSat system, which includes an electrical power system, batteries, an on-board computer and communication systems. A software design tool offers mission design, simulation and analysis, including a link to the hardware for in-loop simulation and testing. A software defined radio link to NANOBED enables ground software validation and operational testing, over which command and control of the system components can be invoked.

The NANOBED EPS provides a 5~V bus suitable for powering the transmitter and receiver FPGA boards directly. For the SPAD supply, the unregulated battery bus was used with a voltage regulator to fix the voltage to 6~V. The total receiver power requirements were 5.37~W, while the transmitter requires 0.905~W. The SPAD consumes the most power in the system, however the commercial module has not been designed with power conservation in mind. Bespoke electronics in place of FPGA boards may also permit lower power consumption, therefore this demonstration should be thought of as an upper limit on power requirements.

\section*{Funding Information}
UK Engineering and Physical Sciences Research Council (EPSRC) EP/M01326X/1, UK Quantum Technology Hub in Quantum Enhanced Imaging.
\\ \\
The NANOBED environment was developed under funding from UK Space Agency in collaboration with Clyde Space Ltd.

\section*{Acknowledgments}

The underlying data for this work can be found at http://dx.doi.org/10.15129/0dfcac12-dbf6-424f-8bc7-71d13db00155
\\ \\
See ``Poissonian communications: optical wireless data transfer at the few-photon level: supplementary material'' for supporting content.

\end{document}